\documentclass[aps,prl,twocolumn,groupedaddress,showpacs]{revtex4}
\usepackage{bm}
\usepackage{graphicx,amssymb}
\usepackage{amsmath}

\newcommand{\beq}{\begin{equation}}
\newcommand{\eeq}{\end{equation}}
\newcommand{\beqa}{\begin{eqnarray}}
\newcommand{\eeqa}{\end{eqnarray}}
\newcommand{\nn}{\nonumber\\}
\renewcommand{\Re}{\text{Re}}

        \newcommand\kk[1]{{{{{#1}}}}}

\begin{document}

\title{Dynamics of a hard sphere granular impurity}
\author{Andr\'es Santos}
\email{andres@unex.es}
\homepage{http://www.unex.es/eweb/fisteor/andres/}
\affiliation{Departamento de F\'{\i}sica, Universidad de
Extremadura, E--06071 Badajoz, Spain}
\author{James W. Dufty}
\email{dufty@phys.ufl.edu}
\homepage{http://www.phys.ufl.edu/~dufty/} \affiliation{Department
of Physics, University of Florida, Gainesville, FL 32611}
\date{\today }

\begin{abstract}
An impurity particle coupling to its host fluid via inelastic hard
sphere collisions is considered. It is shown that the \kk{exact
equation} for its distribution function can be mapped onto that for
an impurity with elastic collisions and an effective mass.
Application of this result to the Enskog--Lorentz kinetic equation
leads to several conclusions: 1) every solution in the elastic case
is equivalent to a class of solutions in the granular case; 2) for
an equilibrium host fluid the granular impurity approaches
equilibrium at a different temperature, with a dominant diffusive
mode at long times; 3) for a granular host fluid in its scaling
state, the granular impurity approaches the corresponding scaling
solution.
\end{abstract}


\maketitle

Granular fluids are the subject of growing attention using a variety
of experimental, numerical, and theoretical tools \cite{JNB96,G03}.
Among the many objectives is a theoretical characterization of
important mechanisms based in the fundamentals of nonequilibrium
statistical mechanics and kinetic theory \cite{G03}. In this
context, an idealized model that captures the single feature of
collisional energy loss has been considered extensively. This is the
system of $N$ smooth, hard spheres with pairwise inelastic
collisions. It has a remarkably rich number of properties that
correlate with those of real granular fluids. Of central interest
are properties that differ from those of normal fluids, and the
reasons behind these differences.

Perhaps the simplest test bed for such questions is a study of the
dynamics of a single impurity particle of mass $m_{0}$ in a
one-component granular fluid. The impurity-fluid particle
interactions have a ``restitution coefficient'' $\alpha _{0}$
measuring the degree of inelasticity ($0<\alpha _{0}\leq 1$, with
$\alpha _{0}=1$ corresponding to elastic collisions). A complete
description of impurity properties is given by the probability
density for its position and velocity at each time,
$F(\mathbf{q}_{0},\mathbf{v}_{0},t)$. Its time dependence is coupled
to the joint distribution for the impurity and a fluid particle,
$f^{\left( 2\right) }\left( x_{0},x_{1},t\right) $, where
$x_{i}\equiv (\mathbf{q}_{i},\mathbf{v}_{i})$, through collisions
between the impurity and fluid particles. This coupling occurs
through an exact equation
\begin{equation}
\left( \partial_t+\mathbf{v}_{0}\cdot {\nabla }_{0}\right)
F(x_{0},t)=L[x_{0};\alpha _{0},m_{0}|f^{(2)}(t)],
\label{1}
\end{equation}
where ${L}[x_{0};\alpha _{0},m_{0}|\cdot ]$ is a linear functional
depending on the impurity position and velocity, and its restitution
coefficient and mass. Equation (\ref{1}) is the exact first
\kk{Bogoliubov--Born--Green--Kirkwood--Yvon (BBGKY)} hierarchy
equation \cite{R77}, with its detailed form given below. A primary
result being reported here is the identity
\begin{equation}
L[x_{0};\alpha _{0},m_{0}|f^{(2)}(t)]=L[x_{0};\alpha
_{0}=1,m_{0}^{\ast }|f^{(2)}(t)],  \label{2}
\end{equation}
where the effective mass is
\begin{equation}
m_{0}^{\ast }=m_{0}+ \left( m_{0}+m\right) \left( 1-\alpha
_{0}\right)/\left( 1+\alpha _{0}\right) , \label{3}
\end{equation}
$m$ being the mass of a fluid particle. This identity establishes an
exact relationship of the coupling of the impurity to its host fluid
by inelastic and elastic collisions. Therefore, it has the potential
to reveal details about the differences and similarities of granular
and normal fluids.

The main utility of (\ref{1}) is its basis for constructing a closed
kinetic equation for $F(x_{0},t)$. In detail, the above functional
$L$ depends on $f^{\left( 2\right) }$ only for the colliding pair at
contact, and on the pre-collision hemisphere. For this
configuration, a kinetic theory results by expressing $f^{\left(
2\right) }\left( x_{0},x_{1},t\right)$
 as a linear functional of $F(x_{0},t)$. A simple and accurate choice is the Enskog--Lorentz
approximation \cite{Dufty}
\begin{equation}
f^{\left( 2\right) }\left( x_{0},x_{1},t\right) \rightarrow
F(x_{0},t) f\left( x_{1},t\right) g\left(
\mathbf{q}_{0},\mathbf{q}_{1}|n\left( t\right) \right) ,  \label{4}
\end{equation}
where $f\left( x_{1},t\right) $ is the fluid probability density and
$g\left( \mathbf{q}_{0},\mathbf{q}_{1}|n\right) $ is the local
equilibrium pair correlation function as a known functional of the
nonequilibrium fluid density $n\left(\mathbf{q}, t\right) $. \kk{The
correlation factor $g$ accounts for the influence of the impurity
particle on the joint impurity-fluid particle distribution. However,
the one-particle fluid distribution $f$ is not affected by the
impurity in the thermodynamic limit ($N\to\infty$).}

Use of the approximation (\ref{4}) in the exact result (\ref{1})
gives the granular Enskog--Lorentz kinetic equation for
$F(x_{0},t)$. The utility of this equation is limited to low or
moderate fluid density and impurity size small compared to the mean
free path. However, within this context, its scope is quite broad,
encompassing homogeneous and inhomogeneous states of the impurity
and arbitrary host fluid states. The identity (\ref{2}) implies that
this kinetic equation is the same as that for an impurity with
elastic collisions but with an effective mass. Several strong
conclusions follow immediately: 1) Since the effective mass is a
function of both the impurity mass and restitution coefficient,
every solution in the elastic case corresponds to a continuous
family of granular solutions with different $\alpha _{0},m_{0}$.
Equivalently, there is a scaling for the granular solutions such
that the dependence on $\alpha _{0},m_{0}$ occurs only through the
single combination $m_{0}^{\ast }$. 2) For the case of a host fluid
at equilibrium, the granular particle approaches equilibrium as
well, but at a different temperature. In addition, the long time
dynamics is hydrodynamic (diffusion) at long wavelengths. 3) For the
case of a host granular fluid in its homogeneous cooling state
\kk{(HCS)}, it is shown below that the impurity particle also
approaches a \kk{HCS} with a different temperature, but the same
cooling rate. These conclusions hold for the complete range of
inelasticity, $0<\alpha _{0}\leq 1$.

There are some limiting cases where some of the implications
discussed here have been noted before. For a massive impurity
particle the Enskog--Lorentz kinetic equation simplifies to a
Fokker--Planck equation, and its exact mapping to the corresponding
elastic case has been given \cite{BDS}. The equilibrium state of an
impurity particle in an equilibrium host fluid has been discussed in
\cite{Martin99}.  It has been noted recently \cite{PVTvW06} that the
impurity mass can be taken as equal to the fluid particle mass by
introducing an effective restitution coefficient, while Piasecki et
al.\ \cite{PTV06} found (considering one-dimensional systems only)
that the special case $m_{0}/m=\alpha _{0}$ is equivalent to $\alpha
_{0}=1$ with $m_{0}^{\ast }/m=1$. All of these special cases are
subsumed by the general results discussed here.

The system considered is a fluid of $N$ hard spheres of mass $m$ and
diameter $\sigma $, and an impurity particle of mass $m_{0}$ and
diameter $\sigma _{0}$. The impurity interacts with the fluid
particles by inelastic hard sphere collisions; the nature of
collisions between a pair of fluid particles is left unspecified at
this point. The reduced distribution function for the impurity obeys
the exact equation (\ref{1}) with the definition
\beqa
L[x_{0};\alpha _{0},m_{0}|f^{(2)}(t)] &\equiv& \overline{\sigma }
^{d-1}\int d\mathbf{v}_{1} d\widehat{\bm{\sigma}}\,\Theta
(\mathbf{v} _{01}\cdot \widehat{\bm{\sigma}})|\mathbf{v}_{01}\cdot
\widehat{\bm{\sigma}} |
                \nn
        &&\times\left[ \alpha _{0}^{-2}f^{\left(
2\right) }( \mathbf{q}_{0},\mathbf{v} _{0}^{\prime \prime
},\mathbf{q}_{0}-\overline{\bm{\sigma}},\mathbf{v} _{1}^{\prime
\prime },t)
    \right.\nn
    &&\left. -f^{\left( 2\right) }(\mathbf{q}_{0},
\mathbf{v}_{0},\mathbf{q}_{0}+\overline{\bm{\sigma}},\mathbf{v}
_{1},t) \right] .
\label{9}
\eeqa
Here, $\overline{\sigma }=\left( \sigma _{0}+\sigma \right) /2$ is
an effective hard sphere diameter for an impurity-fluid pair,
$\overline{\bm{\sigma}}\equiv\overline{\sigma
}\widehat{\bm{\sigma}}$, $d$ is the dimensionality of the system,
$\Theta $ is the Heaviside step function,
$\mathbf{v}_{01}=\mathbf{v}_{0}-\mathbf{v}_{1}$ is the relative
velocity, and
\begin{eqnarray}
\mathbf{v}_{0}^{\prime \prime }
&=&\mathbf{v}_{0}-\frac{m}{m_{0}+m}\frac{1+\alpha _{0}}{\alpha
_{0}}(\mathbf{v}_{01}\cdot \widehat{\bm{\sigma}})
\widehat{\bm{\sigma}},  \notag \\
\mathbf{v}_{1}^{\prime \prime }
&=&\mathbf{v}_{1}+\frac{m_{0}}{m_{0}+m}\frac{ 1+\alpha _{0}}{\alpha
_{0}}(\mathbf{v}_{01}\cdot \widehat{\bm{\sigma}})
\widehat{\bm{\sigma}} \label{10}
\end{eqnarray}
are the restituting collision values  that lead to $\left\{
\mathbf{v}_{0},\mathbf{v}_{1}\right\} $ following a binary
collision.  It is easily verified that the total momentum of the
colliding pair is conserved. Now make the change of variables
$\mathbf{w}_{1}=\mathbf{v}_{1}-\left(\alpha _{0}^{-1}-1\right)
(\mathbf{v}_{01}\cdot \widehat{\bm{\sigma}})\widehat{\bm{\sigma}}$
in the first term of the integrand in (\ref{9}), using
$d\mathbf{v}_{1}=\alpha _{0}d\mathbf{w}_{1}$, $\mathbf{v}_{01}\cdot
\widehat{\bm{\sigma}}=\alpha _{0}\mathbf{w}_{01}\cdot
\widehat{\bm{\sigma}}$, and $\mathbf{w}_{01}\equiv
\mathbf{v}_{0}-\mathbf{w}_{1}$. After some straightforward algebra,
the above scattering law becomes that for an \textit{elastic}
relationship between the pair
$\left(\mathbf{v}_{0},\mathbf{w}_{1}\right)$  and
$\left(\mathbf{v}_{0}^{\prime \prime }, \mathbf{v}_{1}^{\prime
\prime }\right) $,
\begin{eqnarray}
\mathbf{v}_{0}^{\prime \prime }
&=&\mathbf{v}_{0}-\frac{2m}{m_{0}^{\ast }+m}(\mathbf{w}_{01}\cdot \widehat{\bm{\sigma}})\widehat{\bm{\sigma}},  \notag \\
\mathbf{v}_{1}^{\prime \prime } &=&\mathbf{w}_{1}+\frac{2m_{0}^{\ast
}}{m_{0}^{\ast }+m}(\mathbf{w}_{01}\cdot
\widehat{\bm{\sigma}})\widehat{\bm{\sigma}},  \label{13}
\end{eqnarray}
where $m_{0}^{\ast }\left( \alpha _{0},m_{0}\right) $ is defined by
(\ref{3}). Note that this does not imply equivalence between
inelastic and elastic scattering laws. However, only the integration
over $\mathbf{v}_{1}$ is required in (\ref{9}) and this is
equivalent to an integration over the new variable $\mathbf{w}_{1}$.
Finally, by renaming the dummy integration variable
$\mathbf{w}_{1}\rightarrow \mathbf{v}_{1}$ the desired result, Eq.\
(\ref{2}) above, is obtained.

There are some caveats to note at this formally exact level. The
joint distribution $f^{(2)}$ cannot be chosen arbitrarily. In fact,
it obeys an independent second hierarchy equation for which the
dependence on $\alpha _{0}$ cannot be removed by this
transformation. Consequently, while the functional on the right side
of (\ref{2}) is that for an elastic impurity, in general its
argument is not. Hence there is not a complete mapping of the
granular impurity problem to one with elastic collisions. However,
in the approximation to be considered next this mapping is complete.

Use of the approximate functional (\ref{4}) in the first hierarchy
equation gives the revised Enskog--Lorentz kinetic equation
\cite{vanB79,Dufty}
\begin{equation}
\left( \partial_t+\mathbf{v}_{0}\cdot {\nabla }_{0}\right)
F(x_{0},t)=I_{E}[x_{0};\alpha _{0},m_{0}|F(t),f(t)].
\label{21}
\end{equation}
The collision operator, $I_{E}$, is a  functional of both the
impurity and fluid particle distribution functions, $F(x_{0},t)$ and
$f(x_{1},t)$, with the notation $I_{E}[x_{0};\alpha
_{0},m_{0}|F(t),f(t)]\equiv L[x_{0};\alpha _{0},m_{0}|F(t)f(t)g]$.
At low density $g\rightarrow 1$ and this becomes the
Boltzmann--Lorentz--Bogoliubov equation. The identity (\ref{2})
gives directly the corresponding property of the collision operator
$I_{E}[x_{0};\alpha _{0},m_{0}|F(t),f(t)]=I_{E}[x_{0};\alpha
_{0}=1,m_{0}^{\ast }|F(t),f(t)].$ Consequently, this implies the
equality of solutions, for any given fluid state, $F(x_{0},t;\alpha
_{0},m_{0})=F(x_{0},t;\alpha _{0}=1,m_{0}^{\ast }).$ Thus, for any
chosen positive value of $m_{0}^{\ast }\left( \alpha
_{0},m_{0}\right) $ there is an equivalence class of solutions for
different pairs $\left( \alpha _{0},m_{0}\right) $. Furthermore, the
solution for each different class is that for an impurity with
elastic collisions. These are the main observations of the present
work for kinetic theory. It is emphasized that they apply for all
solutions, including time dependent and spatially inhomogeneous
states, and for arbitrary nonequilibrium fluid states as well
\cite{note}. In the following, some of the consequences are noted
for illustration.

For simplicity, the rest of the discussion is limited to cases for
which the host fluid is in a spatially homogeneous state,
$f(\mathbf{q}_{1},\mathbf{v}_{1},t)\rightarrow f(\mathbf{v}_{1},t)$.
Then, evaluated at contact, $g\left(
\mathbf{q}_{0},\mathbf{q}_{1}|n\left( t\right) \right)$ becomes an
overall constant function $\chi(n)$ of the time independent global
density $n=N/V$. Aside from this factor, the collision operator
becomes the usual Boltzmann--Lorentz operator for an impurity of
mass $m_{0}^{\ast }$, i.e., $I_E[x_0;\alpha_0=1,m_0^*|F(t),f(t)]\to
\chi(n)I_B[\mathbf{v}_0|F(t),f(t)]$. Since the equation is linear it
is sufficient to consider a single Fourier mode and the kinetic
equation becomes
\begin{equation}
\partial_t\widetilde{F}(\mathbf{k},\mathbf{v},t)=-i\mathbf{k}\cdot
\mathbf{v}\widetilde{F}(\mathbf{k},\mathbf{v},t)+\chi
(n)I_{B}[\mathbf{v}|\widetilde{F}\left( t\right),f(t) ].
\label{26}
\end{equation}
This is the equation for an impurity with elastic collisions, all
effects of the impurity inelasticity being contained in the
effective mass $m_{0}^{\ast }$.

The condition for a steady state $F(\mathbf{v},t)\rightarrow
F_{e}(\mathbf{v})$ is $\mathbf{k=0}$ and
$I_{B}[\mathbf{v}|F_{e},f(t)]=0$. For an isolated system (e.g.,
unbounded, or periodic boundary conditions) the only solution is a
Maxwellian for both the fluid and impurity distributions, at the
same temperature,
\begin{eqnarray}
f(\mathbf{v},t) &\rightarrow &f_{e}(\mathbf{v})=n\left( {m}/{2\pi T}\right) ^{d/2}e^{-mv^{2}/2T},  \notag \\
F_{e}(\mathbf{v}) &=&\left({m_{0}^{\ast }}/{2\pi T_{0}^{\ast
}}\right) ^{d/2}e^{-m_{0}^{\ast }v^{2}/2T_{0}^{\ast }},  \label{29}
\end{eqnarray}
with $T_{0}^{\ast }=T$. The distribution for the impurity particle
can be expressed in terms of its actual mass $m_{0}$ by introducing
the true impurity temperature $T_{0}\equiv m_{0}T_{0}^{\ast
}/m_{0}^{\ast }=T(1+\alpha_0)/[2+(1-\alpha_0)m/m_0]$. Thus, the
impurity particle is at equilibrium (Maxwellian), just as the fluid
particles, but at a lower temperature $T_{0}<T$. This agrees with
previous results \cite{Martin99} for the stationary state of an
impurity in an equilibrium fluid.

Next, consider the fluid state to be in equilibrium, $f=f_{e}$, and
study the approach of the impurity to its equilibrium distribution
from a general initial state. For this purpose write
\begin{equation}
F(\mathbf{r},\mathbf{v},t)=F_{e}(\mathbf{v})\left[ 1+\phi
(\mathbf{r},\mathbf{v},t)\right] , \label{31}
\end{equation}
so the Enskog--Lorentz equation becomes (in Fourier space)
\begin{equation}
\partial_t\widetilde{\phi} (\mathbf{k,v},t)=\left(
-i\mathbf{k}\cdot \mathbf{v}+\overline{\mathcal{I}}_e\right)
\widetilde{\phi} (\mathbf{k,v},t),
\label{32a}
\end{equation}
where $\overline{\mathcal{I}}_e X=\chi(n)
F_{e}^{-1}I_{B}[\mathbf{v}|F_{e}X,f_e]$. The spectrum of
$-i\mathbf{k}\cdot \mathbf{v}+\overline{\mathcal{I}}_e$ has been
characterized rigorously in the following sense \cite{McL}: The
spectrum of $\overline{\mathcal{I}}_e$ is negative with an isolated
point at zero (conservation of probability). The spectrum of
$-i\mathbf{k}\cdot \mathbf{v}+\overline{\mathcal{I}}_e $ is analytic
in $\mathbf{k}$ about $\mathbf{k=0}$, and its real part is negative.
Hence the point at zero becomes one at $-D(k)$ and remains isolated
from the rest of the spectrum for sufficiently small $k$. In that
limit $D(k)\rightarrow Dk^{2}$ and represents a diffusion mode. In
summary, an initial state decays rapidly to a dominant diffusion
mode and subsequently approaches uniform equilibrium by relaxation
of this hydrodynamic mode. This is the classical picture of ``aging
to hydrodynamics'', followed by approach to equilibrium. The
significance of this well-established result is that it now applies
as well to a granular impurity, through only a change in the
effective mass.

If the host fluid itself is composed of granular particles with
inelastic collisions
then the homogeneous state is no longer equilibrium but
rather the \kk{HCS} \cite{G03,H83}
\begin{equation}
f(\mathbf{v},t)\rightarrow nc^{-d}(t)f_{h}\left( v/c(t)\right)
,\quad c(t)=\sqrt{2T(t)/m},  \label{37}
\end{equation}
where $T(t)$ is related to the average kinetic energy in the usual
way. Its time dependence is due to the continual collisional energy
loss, or ``cooling''. A similar scaling solution for the impurity
particle is sought of the form
\begin{equation}
F(\mathbf{v},t)\rightarrow c_{0}^{-d}(t)F_{h}\left(
v/c_{0}(t)\right) ,\quad c_{0}(t)=\sqrt{2T_{0}^{\ast
}(t)/m_{0}^{\ast }}.  \label{38}
\end{equation}
Here $T_{0}^{\ast }(t)$ is the corresponding measure of the impurity
particle's average kinetic energy. Its time dependence is determined
from the Enskog--Lorentz equation with an explicit time dependence
through the host fluid temperature. If the impurity and fluid
velocities are scaled with $c_{0}(t)$ and $c(t)$, respectively, and
a dimensionless time $s$ is defined through $ds=[c_{0}(t)/\ell]dt$,
with $\ell\equiv 1/\chi(n) n{\overline{\sigma}}^{d-1}$, then the
Enskog--Lorentz collision operator becomes a function of time only
through $T_{0}^{\ast }(t)/T(t)$. The scaling solution (\ref{38})
exists, therefore, only if this ratio is constant, $T_{0}^{\ast
}(t)/T(t)=\gamma $. This does not imply that the temperatures are
the same, only that the cooling rates for the fluid and impurity
particle are the same.

The specific form for $F_{h}\left( v/c_{0}(t)\right)$\kk{, which is
in general different from that of $f_h(v/c(t))$,} is determined from
the Enskog--Lorentz equation. In the above dimensionless units, with
the cooling rate denoted by $\zeta _{0}=-\partial _{s}\ln
T_{0}^{\ast }\left( t\right) $, it becomes the time independent
stationary equation
\begin{equation}
\mathcal{L}F_{h}=0,\hspace{0.3in}\mathcal{L}\equiv -\frac{\zeta
_{0}}{2}\nabla _{\mathbf{v}}\cdot \mathbf{v}+\mathcal{I}.
\label{43}
\end{equation}
Here,  $\mathcal{I}X=I_B[\mathbf{v}|X,f_h]$, except that the
dimensionless relative velocity is
$\mathbf{v}_{01}=\mathbf{v}_{0}-\mathbf{v}_{1}\sqrt{m_0^*/m\gamma}$
and the dimensionless restituting velocities are
$\mathbf{v}_{0}^{\prime \prime }=\mathbf{v}_{0}-\left[ 2m/\left(
m_{0}^{\ast }+m\right) \right] (\mathbf{v}_{01}\cdot
\widehat{\bm{\sigma}})\widehat{\bm{\sigma}}$ and
$\mathbf{v}_{1}^{\prime \prime }=\mathbf{v}_{1}+\left[
2\sqrt{m_{0}^{\ast }m \gamma}/\left( m_{0}^{\ast }+m\right) \right]
(\mathbf{v}_{01}\cdot \widehat{\bm{\sigma}})\widehat{\bm{\sigma}}$.
In this context, it might be expected that initial states for the
impurity approach this stationary HCS solution, analogous to the
approach to equilibrium described above. \kk{Let us prove that this
is indeed the case.}

Paralleling the above analysis, define
\begin{equation}
{F}(\mathbf{r},\mathbf{v},t)=c_{0}^{-d}(t)F_{h}({v}/c_{0}(t))\left[
1+{\phi }(\mathbf{r}/\ell,\mathbf{v}/c_{0}(t),t)\right] ,
\label{46}
\end{equation}
so that for a single Fourier component one has
\begin{equation}
\partial _{s}\widetilde{\phi }=\left( -i\mathbf{k}\cdot
\mathbf{v}+\overline{\mathcal{L}}\right) \widetilde{\phi },\quad
\overline{\mathcal{L}}=-\frac{\zeta _{0}}{2}\mathbf{v}\cdot
\nabla_{\mathbf{v}}-\left(\overline{\mathcal{I}}1\right)
+\overline{\mathcal{I}},
\label{46a}
\end{equation}
with $\overline{\mathcal{L}}\equiv F_{h}^{-1}\mathcal{L}F_{h}$ and
$\overline{\mathcal{I}}\equiv F_{h}^{-1}\mathcal{I}F_{h}$. Define a
Hilbert space with scalar product $\left( a,b\right) \equiv \int
d\mathbf{v}F_{h}\left( v\right) a^{+}\left(\mathbf{v}\right)
b\left(\mathbf{v}\right)$. Then, the following identity applies
\begin{equation}
\Re \left( \psi ,\overline{\mathcal{L}}\psi \right) =\Re \left( \psi
,\overline{\mathcal{I}}\psi \right) -\frac{1}{2}\left( \psi
,\psi\overline{\mathcal{I}}1 \right)  .
\label{48}
\end{equation}
Next, note the representation
\begin{eqnarray}
\left( \varphi ,\overline{\mathcal{I}}\psi \right)  &=&\int
d\mathbf{v}_{0}d\mathbf{v}_{1}d\widehat{\bm{\sigma}}F_{h}(\mathbf{v}_{0})f_{h}(\mathbf{v}
_{1})\Theta (\mathbf{v}_{01}\cdot \widehat{\bm{\sigma}})|\mathbf{v}
_{01}\cdot \widehat{\bm{\sigma}}|  \notag \\
&&\times \left[ \varphi ^{+}\left( \mathbf{v}_{0}^{\prime \prime
}\right) -\varphi ^{+}\left( \mathbf{v}_{0}\right) \right] \psi
\left( \mathbf{v}_{0}\right) ,  \label{49}
\end{eqnarray}
where use has been made of the consequence of elastic collisions
$d\mathbf{v}_{0}d\mathbf{v}_{1}=d\mathbf{v}_{0}^{\prime \prime
}d\mathbf{v}_{1}^{\prime \prime }$, $\mathbf{v}_{01}\cdot
\widehat{\bm{\sigma}}=-\mathbf{v}_{01}^{\prime \prime }\cdot
\widehat{\bm{\sigma}}$, and the fact that the inverse and direct
scattering laws are the same. Applying this to the right side of
(\ref{48}) gives the desired result
\begin{equation}
\Re \left( \psi ,\left( -i\mathbf{k}\cdot
\mathbf{v}+\overline{\mathcal{L}}\right) \psi \right) \leq 0,
\label{14}
\end{equation}
i.e., the generator for the dynamics is dissipative \cite{Scharf67}.
A similar analysis of $\left( \varphi ,\overline{\mathcal{L}}\psi
\right) $ shows that a constant is an eigenfunction of both
$\overline{\mathcal{L}}$ and its adjoint, with zero eigenvalue. The
null space is therefore spanned by $1$. By normalization the
solution $\widetilde{\phi } $ in (\ref{46a}) is orthogonal to $1$.
Consequently,
\begin{equation}
\partial _{s}\| \widetilde{\phi }\| ^{2}=2\Re \left(
\widetilde{\phi },\partial _{s}\widetilde{\phi }\right) <0.
\label{16}
\end{equation}
Since the inequality is strict now, $\| \widetilde{\phi }\| $ is
positive, monotonically decreasing, and must go to zero.

In summary, the connection (\ref{2}) obtained here presents the
possibility to explore new consequences for the kinetic theory of
granular impurity dynamics. This has been illustrated for the case
of the Enskog--Lorentz kinetic equation. The result (\ref{32a}) for
the equilibrium host fluid is significant since it establishes the
fact that inelasticity does not affect any qualitative features of
the impurity dynamics in that case. In particular, the existence of
a diffusion mode, its dominance  at long times, and approach to
equilibrium still apply. In the more interesting case of a granular
host fluid, the result (\ref{16}) is significant in showing that the
non-trivial scaling solution $F_{h}$ is universal in the same sense
as the Maxwellian for normal fluids. Further study of the
inhomogeneous generator $-i\mathbf{k}\cdot
\mathbf{v}+\overline{\mathcal{L}}$ is needed to answer such
questions as: do spatial perturbations exhibit a diffusive mode?; if
so, does it exist and dominate at long times even for strong
dissipation? Formally, the diffusion mode can be identified by
assuming spectral isolation and analyticity in $\mathbf{k}$ for a
perturbation expansion of the zero eigenvalue of
$\overline{\mathcal{L}}$ \cite{DB05}. However, analysis in the large
mass limit suggests that the answers may be conditional. In that
limit it has been shown \cite{BDS} that diffusion requires that the
cooling rate of the host fluid not be too large compared to the
impurity-fluid collision rate. More complete characterization of the
spectrum of the generators of dynamics for inhomogeneous states
seems to be limited at present to idealized models \cite{DB05,BD05}.

Kinetic theory for granular fluids and its domain of validity remain
topics of intense debate. In practice, limitations due to
approximate implementations can become confused with those of the
theory itself. In this context, precise results for restricted
systems (e.g., an impurity particle) can provide instructive
benchmarks for important conceptual issues. This is one of the
objectives of the present analysis.

The research of A.S. has been partial supported by the Ministerio de
Educaci\'on y Ciencia (Spain) through grant No.\ FIS2004-01399
(partially financed by FEDER funds).


\begin{thebibliography}{99}
\bibitem{note}
\kk{Note that the identity (\protect\ref{2}) and  the results
reported in this Letter still hold if the collision kernel is
modified as, for instance, in the inelastic Maxwell model.}
\end{thebibliography}
\end{document}